\pdfoutput = 1
\documentclass[letterpaper, conference, twocolumn]{IEEEtran}
\usepackage{cite}
\usepackage{amsmath}
\usepackage{array}
\ifCLASSOPTIONcompsoc
  \usepackage[caption=false,font=normalsize,labelfont=sf,textfont=sf]{subfig}
\else
  \usepackage[caption=false,font=footnotesize]{subfig}
\fi
\usepackage{url}

\usepackage{amssymb}
\usepackage{nth}
\usepackage{bm}
\usepackage{adjustbox}
\usepackage{siunitx}
\usepackage{xcolor}
\usepackage{graphicx}
\usepackage{enumerate}

\usepackage{tikz}

\usetikzlibrary{backgrounds, calc, decorations, decorations.markings, decorations.pathreplacing, math, arrows.meta, shapes, patterns}
\pgfdeclarelayer{bg}    
\pgfdeclarelayer{bg2}    
\pgfsetlayers{bg2,bg,main}  

\usepackage{pgfplots}
\usepgfplotslibrary{colormaps}

\makeatletter
\renewcommand*\env@matrix[1][c]{\hskip -\arraycolsep
	\let\@ifnextchar\new@ifnextchar
	\array{*\c@MaxMatrixCols #1}}
\makeatother

\newcommand{\vect}[1]{\bm{#1}}

\newcommand{\trans}{^{\mathrm{T}}}
\newcommand{\herm}{^{\mathrm{H}}}
\newcommand{\e}{\mathrm{e}}

\DeclareMathOperator{\jj}{j}

\DeclareMathOperator*{\atan2}{atan2}

\title{Multi-Array 5G V2V Relative Positioning: Performance Bounds}

\author{\IEEEauthorblockN{Anastasios Kakkavas\IEEEauthorrefmark{1}\IEEEauthorrefmark{2}, Mario H. Casta\~neda Garc\'ia\IEEEauthorrefmark{1}, Richard A. Stirling-Gallacher\IEEEauthorrefmark{1} and Josef A. Nossek\IEEEauthorrefmark{2}\IEEEauthorrefmark{3}}
	\IEEEauthorblockA{\IEEEauthorrefmark{1}Munich Research Center, Huawei Technologies D\"usseldorf GmbH, Munich, Germany}
	\IEEEauthorblockA{\IEEEauthorrefmark{2}Department of Electrical and Computer Engineering, Technische Universit{\"{a}}t M{\"{u}}nchen, Munich, Germany}
	\IEEEauthorblockA{\IEEEauthorrefmark{3}Department of Teleinformatics Engineering, Federal University of Cear{\'a}, Fortaleza, Brazil}
	\IEEEauthorblockA{
		\texttt{\{anastasios.kakkavas, mario.castaneda, richard.sg\}@huawei.com, josef.a.nossek@tum.de}
	}
}
\usepackage{textcomp}
\newcommand\copyrighttext{%
	\footnotesize \textcopyright 2018 IEEE. Personal use of this material is permitted. Permission from IEEE must be obtained for all other uses, in any current or future media, including reprinting/republishing this material for advertising or promotional purposes, creating new collective works, for resale or redistribution to servers or lists, or reuse of any copyrighted component of this work in other works.}

\newcommand\copyrightnotice{%
	\begin{tikzpicture}[remember picture,overlay]
	\node[anchor=south,yshift=10pt] at (current page.south) {{\parbox{\dimexpr\textwidth-\fboxsep-\fboxrule\relax}{\copyrighttext}}};
	\end{tikzpicture}%
}

\newcommand\conferenceinfotext{%
	\footnotesize A. Kakkavas, M. H. Casta\~neda Garc\'ia, R.A. Strirling-Gallacher and Josef A. Nossek, “Multi-Array 5G V2V Relative Positioning: Performance Bounds,” in GLOBECOM 2018 - 2018 IEEE
	Global Communications Conference, Abu Dhabi, U.A.E., Dec. 2018, pp. 206-212. doi: 10.1109/GLOCOM.2018.8647812}

\newcommand\conferenceinfonotice{%
	\begin{tikzpicture}[remember picture,overlay]
	\node[anchor=south,yshift=-27pt] at (current page.north) {{\parbox{\dimexpr\textwidth-\fboxsep-\fboxrule\relax}{\conferenceinfotext}}};
	\end{tikzpicture}%
}

\usepackage{hyperref}
\hypersetup{
	pdfinfo={
		Title={Multi-Array 5G V2V Relative Positioning: Performance Bounds},
		Author={Anastasios Kakkavas, Mario H. Casta\~neda Garc\'ia, Richard A. Strirling-Gallacher and Josef A. Nossek}
	}
}

\begin{document}
\maketitle
\IEEEoverridecommandlockouts
\copyrightnotice
\conferenceinfonotice

\vspace*{-0.35cm}
\begin{abstract}
	We study the performance bounds of vehicle-to-vehicle (V2V) relative positioning for vehicles with multiple antenna arrays.
	The Cram\'{e}r-Rao bound for the estimation of the relative position and the orientation of the Tx vehicle is derived, when angle of arrival (AOA) measurements with or without time-difference of arrival (TDOA) measurements are used. In addition, geometrically intuitive expressions for the corresponding Fisher information are provided. The derived bounds are numerically evaluated for different carrier frequencies, bandwidths and array configurations under different V2V scenarios, i.e. overtaking and platooning. The significance of the AOA and TDOA measurements for position estimation is investigated. The achievable positioning accuracy is then compared with the present requirements of  the 3rd Generation Partnership Project (3GPP) 5G New Radio (NR) vehicle-to-everything (V2X) 
	standardization.
\end{abstract}

\section{Introduction}
	Positioning is expected to have an upgraded role in the fifth generation (5G) of wireless networks, compared to its predecessors, as position information will not only be offered as a service, but will also be used as an aid for communication-related tasks, such as proactive resource allocation and beamforming~\cite{HWC+15}. Massive number of antennas, large bandwidth available at millimeter-Wave frequencies and dense base station deployment are considered as key enablers not only for high data rates, but also for accurate position estimation~\cite{SGD+18}. The improved positioning capability of 5G networks can prove extremely useful in a wide variety of scenarios, like assisted living~\cite{WML+16}, smart factories~\cite{WHK+16} and automated driving~\cite{WSD+17}, where existing positioning technologies, such as Global Navigation Satellite Systems (GNSS), are not able to guarantee the necessary positioning accuracy under all conditions.
	
	An important use case in the context of automated driving is user equipment (UE)-centric vehicle-to-vehicle (V2V) relative positioning~\cite{HKC+18}, which is the focus of this work. Several works, directly or indirectly relevant to the V2V positioning scenario, have been published. The performance limits of multi-anchor array localization using time of arrival (TOA) and angle of arrival (AOA) measurements were studied in~\cite{HSZ+16}. In~\cite{SGD+15} the Cram\'{e}r-Rao bound (CRB) for single-anchor line-of-sight (LOS) receiver (Rx) position estimation was presented, when angle of departure (AOD), AOA and TOA measurements are available. 
	Single-anchor transmitter (Tx) localization error bounds for different array configurations 
	were studied in \cite{GGD17}, taking synchronization and quantization errors of beamforming weights into account. In \cite{AZA+17} asymptotic expressions for the position error bounds of systems with large bandwidth and large number of antennas were derived.
	Building on this work, the authors of \cite{MWB+17} showed that the Fisher information provided by single-bounce non-LOS (NLOS) paths is rank-$1$, while also deriving analytic expressions for the eigendirection and eigenvalue of the information. In \cite{SGD+18}, the single-anchor Rx position and orientation error bounds were studied, when a multicarrier waveform is used, and an algorithm which approaches these bounds in the high signal-to-noise ratio (SNR) regime was presented. Apart from general studies on positioning, which are also applicable to the V2V scenario, some V2V positioning-specific works have been recently published. A setup with multiple on-vehicle Rx arrays is considered in \cite{DDG+17} and the TOA-based position error bound is derived. 
	A method for sensing vehicles with multiple Tx arrays under NLOS-only propagation was presented in \cite{HKC+18}.
	
	In this work we consider Tx and Rx vehicles equipped with multiple antenna arrays, referred to as panels in 5G New Radio
	(NR) standardization, which are distributed around the vehicle to support 5G NR side link (V2V) communication between vehicles. Our goal is to leverage these arrays and the side link to also perform V2V relative positioning using e.g. position reference signals. 
	We assume the clocks between the Tx and Rx vehicles are not synchronized or the time of transmission is not known. Thus, the TOA cannot be computed and only time-difference of arrival (TDOA) measurements are possible. We consider short reference signal transmission time and small relative velocity between vehicles, such that the setup is static over the observation interval. 
	To this end, we avoid a beam-scanning procedure at the Tx and assume the reference signal is transmitted through a fixed (wide) beam. Consequently, angles of departure (AODs) cannot be measured; thus, they are not considered for position estimation.
	Being interested in relatively short distance between neighbor vehicles, which is mainly dominated by LOS paths, we only consider LOS propagation for simplicity and to obtain initial insight. Assuming the Tx arrays have orthogonal frequency division multiple access (OFDMA) to the shared medium, we derive geometrically intuitive expressions for the Fisher information on the relative position and orientation of the Tx vehicle, 
	when AOA measurements with or without TDOA measurements are available. These expressions are then used to obtain the CRB for the lateral and longitudinal positioning error, which are evaluated for various scenarios and system configurations,
	to determine the range of distances for which the requirements set by 5G NR V2X standardization \cite{TS22.186} are met.
	We note that, since AODs cannot be observed, the Tx-Rx pairs with NLOS-only links, which are ignored here, do not offer positioning information~\cite{HSZ+16,MWB+17}. 
	
	The rest of the paper is organized as follows. The system model is derived in Section II and the Cram\'{e}r-Rao bound for relative position and orientation estimation is derived in Section III. Numerical evaluations of the bounds are provided in Section IV and Section V concludes the work.

\section{System Model}
\begin{figure}
		\centering
		\begin{adjustbox}{scale=0.87}\includegraphics[]{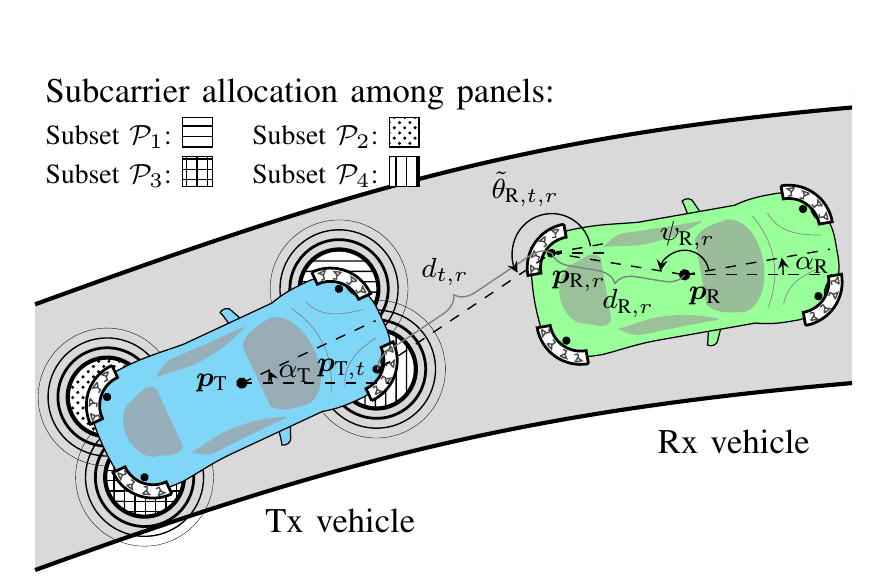}\end{adjustbox}
		\caption{Geometric model of V2V multi-array scenario and illustration of the resource allocation among panels.}
		\label{fig:geometric model} 
	\end{figure}

	In Fig. \ref{fig:geometric model} the considered setup is depicted. 
	The position of the Tx and Rx vehicles is given by the reference points $\vect{p}_{\text{T}}$ and $\vect{p}_{\text{R}} = \vect{p}_{\text{T}} + \vect{q}\in \mathbb{R}^2$, where $\vect{q} = [q_{\text{x}}, q_{\text{y}}]\trans$ is the relative position of the Rx vehicle with respect to (w.r.t.) the Tx vehicle, with $(\cdot)\trans$ denoting transposition, and the orientation $\alpha_{\text{T}}$ and $\alpha_{\text{R}}$ of the vehicles is given w.r.t. 
	the horizontal half-line originating from the vehicle's reference point (see Fig. \ref{fig:geometric model}).
	The Tx and Rx vehicles are equipped with $K_{\text{T}}$ and $K_{\text{R}}$ antenna arrays, with the $t$-th Tx array having $N_{\text{T},t}$ antennas and the $r$-th Rx array having $N_{\text{R},r}$ antennas, respectively.
	Setting $\vect{u}\left(\psi\right) = \left[\cos(\psi),\;\sin(\psi)\right]\trans$, 
	the centroid of the $r$-th Rx array is at $\vect{p}_{\text{R},r} = \vect{p}_{\text{R}} + d_{\text{R},r}\vect{u}\left(\psi_{\text{R},r} + \alpha_{\text{R}}\right) = [p_{\text{R},r,\text{x}},\; p_{\text{R},r,\text{y}}]\trans,r=1,\ldots,K_{\text{R}}$, 
	where $d_{\text{R},r}$ and $\psi_{\text{R},r} + \alpha_{\text{R}}$ are its distance and angle from the reference point of the Rx vehicle, as shown in Fig. \ref{fig:geometric model}. 
	The $i$-th element of the $r$-th Rx array is located at $\vect{p}_{\text{R},r,i} = \vect{p}_{\text{R},r} + d_{\text{R},r,i}\vect{u}\left(\psi_{\text{R},r,i} + \alpha_{\text{R}}\right),i=1,\ldots,N_{\text{R},r}$, with $d_{\text{R},r,i}$ and $\psi_{\text{R},r,i} + \alpha_{\text{R}}$ being its distance and angle from the array's centroid. 
	The quantities for the Tx arrays are defined similarly. 
	We note here that the optimization of the arrays' position is outside the scope of this work.
	
	We consider an OFDM system with $N$ subcarriers and symbol duration (without the cyclic prefix) $NT_{\text{s}}$, where $T_{\text{s}}$ is the sample period and $F_{\text{s}} = 1/T_{\text{s}}$ is the sampling rate. The $t$-th Tx array uses a subset $\mathcal{P}_t$ of all the occupied subcarriers $\mathcal{P} = \cup_{t=1}^{K_{\text{T}}}\mathcal{P}_t$, with $|\mathcal{P}|\leq N$ and $\mathcal{P}_t\cap\mathcal{P}_{t'}=\varnothing$ for $t\neq t'$, so that the Tx arrays have orthogonal access to the channel (see Fig. \ref{fig:geometric model}). We note that orthogonal access could also be enabled over time or code.
	The relative velocity between the vehicles is such that the channel is static during the transmission of the reference signal. 
	We consider a V2V scenario with short distances and a pure LOS connection between neighbor vehicles, assuming that for such distances the NLOS components of the channel are negligible.
	However, there is not always a LOS between all Tx-Rx array pairs, depending on the position and orientation of the vehicles.
	Assuming the aperture of the arrays is much smaller than the distance of the Tx and Rx arrays, 
	the propagation delay from the $j$-th antenna of the $t$-th Tx panel to the $i$-th antenna of the $r$-th Rx panel is well approximated by $\tau_{t,r,j,i} \approx \tau_{t,r} - \tau_{\text{T},t,j}(\tilde{\theta}_{\text{T},t,r}) - \tau_{\text{R},r,i}(\tilde{\theta}_{\text{R},t,r})$, where 
	\begin{IEEEeqnarray}{rCl}
		\tau_{t,r} &=& d_{t,r}/c, \\
		\tau_{\text{T},t,j}(\tilde{\theta}_{\text{T},t,r}) &=& d_{\text{T},t,j}\vect{u}\trans(\psi_{\text{T},t,j}) \vect{u}(\tilde{\theta}_{\text{T},t,r})/c\\
		\tau_{\text{R},r,i}(\tilde{\theta}_{\text{R},t,r}) &=& d_{\text{R},r,i}\vect{u}\trans(\psi_{\text{R},r,i}) \vect{u}(\tilde{\theta}_{R,l})/c,
	\end{IEEEeqnarray} 
	with $d_{t,r} = \left\|\vect{p}_{\text{R},r} - \vect{p}_{\text{T},t}\right\|_2$, $\tilde{\theta}_{\text{T},t,r} = \theta_{\text{T},t,r} - \alpha_{\text{T}}, \tilde{\theta}_{\text{R},t,r} = \theta_{\text{R},t,r} - \alpha_{\text{R}}, \theta_{\text{T},t,r} = \theta_{\text{R},t,r} + \pi, \theta_{\text{R},t,r} = \atan2(p_{\text{R},r,y} - p_{\text{T},t,y}, p_{\text{R},r,x} - p_{\text{T},t,x})$, and $\atan2(y,x)$ being the four-quadrant inverse tangent function. Without loss of generality we assume that $\tau_{1,1} = \min_{t,r} \tau_{t,r}$.
	The reference signal from each Tx array is transmitted through a fixed beamforming vector $\vect{f}_t\in\mathbb{C}^{N_{\text{T},t}}$ over $N_{\text{B}}$ OFDM symbols. We assume a properly designed cyclic prefix is available and coarse synchronization is such that inter-block interference is avoided. The received signal at the $r$-th Rx array at subcarrier $p\in\mathcal{P}_t$ of the $b$-th OFDM symbol, with $b=1,\ldots,N_{\text{B}}$, is
	\begin{IEEEeqnarray}{rCl}
		\vect{y}_{r,b}[p] &=& \e^{-\jj \omega_p(\tau_{\text{s}} + \Delta\tau_{t,r})} h_{t,r}'\vect{a}_{\text{R},r}(\tilde{\theta}_{\text{R},t,r})\times\nonumber\\
		&&\quad \vect{a}_{\text{T},t}\trans(\tilde{\theta}_{\text{T},t,r})\vect{f}_{t}x_b[p] + \vect{\eta}_{r,b}[p]\nonumber\\
		&=& \e^{-\jj \omega_p(\tau_{\text{s}} + \Delta\tau_{t,r})} h_{t,r}\vect{a}_{\text{R},r}(\tilde{\theta}_{\text{R},t,r})x_b[p] + \vect{\eta}_{r,b}[p],
		\label{eq:signal model}
		\IEEEeqnarraynumspace
	\end{IEEEeqnarray}
	where $x_b[p]$ is the transmit symbol at the $p$-th subcarrier of the $b$-th OFDM symbol, $\omega_p = 2 \pi p F_{\text{s}}/N, \;\tau_{\text{s}}$ is the timing offset of the coarse synchronization from the TOA of the shortest path, $\vect{\eta}_{r,b}[p]$ is the spatially and temporally white Gaussian noise with per-antenna variance $\sigma_{\eta,r}^2$, $h_{t,r}'$ is the complex gain of the link between the $t$-th Tx and the $r$-th Rx array, and
	\begin{IEEEeqnarray}{rCl}
		\vect{a}_{\text{T},t}(\tilde{\theta}_{\text{T},t,r}) &=& \Big[e^{\jj \omega_{\text{c}} \tau_{\text{T},t,1}(\tilde{\theta}_{\text{T},t,r})}, \ldots, \e^{\jj \omega_{\text{c}} \tau_{\text{T},t,N_{\text{T},t}}(\tilde{\theta}_{\text{T},t,r})}\Big]
		\trans\label{eq:tx steering vector}\\
		\vect{a}_{\text{R},r}(\tilde{\theta}_{\text{R},t,r}) &=& \Big[e^{\jj \omega_{\text{c}} \tau_{\text{R},r,1}(\tilde{\theta}_{\text{R},t,r})}, \ldots, \e^{\jj \omega_{\text{c}} \tau_{\text{R},r,N_{\text{R},r}}(\tilde{\theta}_{\text{R},t,r})}\Big]
		\trans\label{eq:rx steering vector}
		\IEEEeqnarraynumspace
	\end{IEEEeqnarray} 
	are the steering vectors of the $t$-th Tx array and the $r$-th Rx array, respectively, 
	where $\omega_{\text{c}}=2\pi f_{\text{c}}$, with $f_{\text{c}}$ as the carrier frequency.
	For the derivation of the signal model \eqref{eq:signal model} we have assumed a narrowband system, i.e. $\omega_{\text{c}} \gg \omega_{p_{\max}}$, where $p_{\max}$ is the maximum subcarrier index in $\mathcal{P}$. We have set $h_{t,r} = h_{t,r}'\vect{a}_{\text{T},t}\trans(\tilde{\theta}_{\text{T},t,r})\vect{f}_{t}$, as, due to the fixed beamforming vector $\vect{f}_t$ and the unknown channel gains, the AOD $\tilde{\theta}_{\text{T},t,r}$ cannot be observed. With $P_{\text{T}}$ being the Tx power per OFDM symbol for all Tx arrays, the power allocated to the $t$-th Tx array is $\gamma_t P_{\text{T}}$, where $\gamma_t\in [0,1]$ and $\sum\nolimits_{t=1}^{K_T}\gamma_t=1$. The power allocated to the $p$-th subcarrier, with $p\in\mathcal{P}_t$, over $N_{\text{B}}$ OFDM symbols is $\sum_{b=0}^{N_{\text{B}} - 1} |x_b[p]|^2 = \gamma_t \gamma_{t,p} N_{\text{B}} P_{\text{T}}$, where $\gamma_{t,p}\in [0,1]$ describes the fraction of power of the $t$-th Tx array allocated to the $p$-th subcarrier, with $\sum_{p\in\mathcal{P}_t} \gamma_{t,p} = 1$. 

\section{Cram\'{e}r-Rao bound for Relative Position and Orientation Estimation}
	\label{sec:CRLB computation}
	Based on the received signal model presented in the last section, we now derive the CRB for the estimation of the relative position $\vect{q}$ and the orientation $\alpha_{\text{T}}$ of the Tx vehicle. 
	Setting $h_{t,r,\Re} = \Re\{h_{t,r}\}$ and $h_{t,r,\Im} = \Im\{h_{t,r}\}, \quad t=1,\ldots,K_{\text{T}}, r=1,\ldots,K_{\text{R}}$, we define the channel parameter vector $\vect{\phi}\in \mathbb{R}^{4K_{\text{T}}K_{\text{R}}}$ as
	\begin{IEEEeqnarray}{rCl}
		\vect{\phi} &=& \hspace*{-0.05cm}[\tau_{\text{s}}, \tilde{\theta}_{\text{R},1,1}, h_{1,1, \Re}, h_{1,1, \Im},\Delta\tau_{1,2}, \tilde{\theta}_{\text{R},1,2}, h_{1,2,\Re}, h_{1,2,\Im}, \ldots, \hspace*{-0.03cm}\nonumber\\
		&&\Delta\tau_{K_{\text{T}},K_{\text{R}}}, \tilde{\theta}_{\text{R},K_{\text{T}},K_{\text{R}}}, h_{K_{\text{T}},K_{\text{R}},\Re}, h_{K_{\text{T}},K_{\text{R}},\Im}
		]\trans.
		\IEEEeqnarraynumspace
	\end{IEEEeqnarray}
	Then, since $\vect{\phi}$ is observed under Gaussian noise, the covariance matrix $\vect{C}_{\hat{\vect{\phi}}}$ of any unbiased estimator $\hat{\vect{\phi}}$ satisfies~\cite{Kay93}
	\begin{IEEEeqnarray}{rCl}
		\vect{C}_{\hat{\vect{\phi}}} - \vect{J}_{\vect{\phi}}^{-1} \succeq \vect{0},
		\IEEEeqnarraynumspace
	\end{IEEEeqnarray}
	where $\succeq \vect{0}$ denotes positive semi-definiteness and $\vect{J}_{\vect{\phi}}$ is the FIM of $\vect{\phi}$.
	The $(i,j)$-th entry of $\vect{J}_{\vect{\phi}}\in\mathbb{R}^{4K_{\text{T}}K_{\text{R}}\times 4K_{\text{T}}K_{\text{R}}}$ is 
	\begin{IEEEeqnarray}{rCl}
	\left[\vect{J}_{\vect{\phi}}\right]_{i,j} \hspace*{-0.02cm}&=&\hspace*{-0.02cm} \sum_{r=1}^{K_{\text{R}}}\hspace*{-0.02cm}\frac{2}{\sigma_{\eta,r}^2}\sum_{b=1}^{N_{\text{B}}}\sum_{t=1}^{K_{\text{T}}}\hspace*{-0.02cm} \sum_{p\in \mathcal{P}_t} \hspace*{-0.05cm}\Re\hspace*{-0.03cm}\left\{\frac{\partial \vect{m}_{r,b}\herm[p]}{\partial \left[\vect{\phi}\right]_i} \frac{\partial \vect{m}_{r,b}[p]}{\partial \left[\vect{\phi}\right]_j}\right\}\hspace*{-0.02cm},
	\label{eq:entries of channel parameter FIM}
	\IEEEeqnarraynumspace
	\end{IEEEeqnarray}
	where
	\begin{IEEEeqnarray}{rCl}
	{\vect{m}}_{r,b}[p]\hspace*{-0.03cm} &=&\hspace*{-0.03cm} \e^{-\hspace*{-0.03cm} \jj\hspace*{-0.03cm} \omega_p(\tau_{\text{s}} \hspace*{-0.005cm} + \hspace*{-0.005cm}\Delta\tau_{t,r})}\hspace*{-0.02cm}  h_{t,r}\vect{a}_{\text{R},r}(\tilde{\theta}_{\text{R},t,r})x_b[p]\hspace*{-0.03cm}\in\hspace*{-0.03cm}\mathbb{C}^{N_{\text{R},r}}\hspace*{-0.03cm} .
	\IEEEeqnarraynumspace
	\end{IEEEeqnarray}
	The required derivatives of ${\vect{m}}_{r,b}[p]$ in \eqref{eq:entries of channel parameter FIM} are
	\begin{IEEEeqnarray}{rCl}
		\frac{\partial \vect{m}_{r,b}[p]}{\partial \tau_{\text{s}}} &=& -\jj \omega_p \vect{m}_{r,b}[p] \\
		\frac{\partial \vect{m}_{r,b}[p]}{\partial \Delta\tau_{t',r'}} &=& \left\{ \begin{array}{@{}ll@{}}
				-\jj \omega_p \vect{m}_{r',b}[p], & r=r', p\in\mathcal{P}_{t'}\\
				0,  & \mbox{otherwise}
			\end{array}
			\right.\\
		\frac{\partial \vect{m}_{r,b}[p]}{\partial \tilde{\theta}_{R,t',r'}} &=& \left\{ \begin{array}{@{}ll@{}}
				\vect{D}_{\text{R},r'}(\tilde{\theta}_{\text{R},t',r'}) \vect{m}_{r',b}[p], & \hspace*{-0.01cm}r=r', p\in\mathcal{P}_{t'}\\
				0,  & \hspace*{-0.01cm}\mbox{otherwise}
			\end{array}
			\right.\hspace*{-1.5pt}\\
		\frac{\partial \vect{m}_{r,b}[p]}{\partial h_{t',r', \Re}} &=& -\jj\frac{\partial \vect{m}_{r,b}[p]}{\partial  h_{t',r', \Im}} = \left\{ \begin{array}{@{}ll@{}}
			\frac{\vect{m}_{r',b}[p]}{h_{t',r'}}, & \hspace*{-0.01cm}r=r', p\in\mathcal{P}_{t'}\\
			0,  & \hspace*{-0.01cm}\mbox{otherwise,}
		\end{array}
		\right.\hspace*{-1.5pt}
		\IEEEeqnarraynumspace
	\end{IEEEeqnarray}
	where $\vect{D}_{\text{R},r'}(\tilde{\theta}_{\text{R},t',r'})\in \mathbb{C}^{N_{\text{R},r'}}$ is a diagonal matrix, with $[\vect{D}_{\text{R},r'}(\tilde{\theta}_{\text{R},t',r'})]_{i,i} = -\jj \omega_{\text{c}} d_{\text{R},r',i}\vect{u}\trans(\psi_{\text{R},r',i}) \vect{u}_{\perp}(\tilde{\theta}_{\text{R},t',r'})/c$ and $\vect{u}_{\perp}\left(\psi\right) = \vect{u}\left(\psi - {\pi}/{2}\right)$.
	As we are interested in the estimation accuracy of the relative position $\vect{q}$ and the Tx vehicle orientation $\alpha_{\text{T}}$, we need to determine how the information on the angles of arrival (AOAs) $\tilde{\theta}_{\text{R},t,r}$ and time-differences of arrival (TDOAs) $\Delta\tau_{t,r}$ is translated to information about $\vect{q}$ and $\alpha_{\text{T}}$. The relation between AOAs and TDOAs and position and orientation parameters is described by
	\begin{IEEEeqnarray}{rCl}
		c\Delta\tau_{t,r} &=& \left\| \vect{q}_{t,r} \right\|_2 - \left\| \vect{q}_{1,1} \right\|_2\\
		\tilde{\theta}_{\text{R},t,r} &=& \atan2\left( \left[0\;\;1\right]\vect{q}_{t,r}, \left[1\;\;0\right]\vect{q}_{t,r} \right) - \alpha_{\text{R}},
		\IEEEeqnarraynumspace
	\end{IEEEeqnarray}
	where
	$\vect{q}_{t,r} = \vect{p}_{\text{T},t} - \vect{p}_{\text{T}} - \vect{p}_{\text{R},r} + \vect{p}_{\text{R}} - \vect{q}$.	
	We consider two cases, where different sets of measurements are used for estimating relative position and orientation: i) Both AOA and TDOA measurements or ii) only AOA measurements are used.
	
	\subsection{Using AOA and TDOA measurements}
		We define the parameter vector containing the position, orientation and nuisance parameters in the model as
		\begin{IEEEeqnarray}{rCl}
			\tilde{\vect{\phi}} &=& [\vect{q}\trans, \alpha_{\text{T}}, \tau_{\text{s}}, h_{1,1, \Re}, h_{1,1, \Im}, \ldots, \nonumber\\
			&& h_{K_{\text{T}},K_{\text{R}}, \Re}, h_{K_{\text{T}},K_{\text{R}}, \Im}
			]\trans\in\mathbb{R}^{4 + 2K_{\text{T}} K_{\text{R}}}\label{eq:position parameter vector AOA_TDOA}.
			\IEEEeqnarraynumspace
		\end{IEEEeqnarray}
		The FIM $\vect{J}_{\tilde{\vect{\phi}}}$ of $\tilde{\vect{\phi}}$ can be obtained using $\vect{J}_{\vect{\phi}}$ and the transformation matrix $\vect{T}\in\mathbb{C}^{(4+2K_{\text{T}} K_{\text{R}})\times 4K_{\text{T}} K_{\text{R}}}$~\cite{LC98}:
		\begin{IEEEeqnarray}{rCl}
			\vect{J}_{\tilde{\vect{\phi}}} &=& \vect{T}\vect{J}_{\vect{\phi}}\vect{T}\trans,
			\IEEEeqnarraynumspace
		\end{IEEEeqnarray}
		where
		\begin{IEEEeqnarray}{rCl}
			\left[\vect{T}\right]_{i,j} &=& \partial \left[\vect{\phi}\right]_j/\partial \left[\tilde{\vect{\phi}}\right]_i.
			\IEEEeqnarraynumspace
		\end{IEEEeqnarray}
		It is trivial that the entries of $\vect{T}$ corresponding to identical parameters in $\vect{\phi}$ and $\tilde{\vect{\phi}}$ are equal to $1$, e.g. $\partial \tau_{\text{s}}/\partial\tau_{\text{s}}=1$. The rest non-zero entries are
		\begin{IEEEeqnarray}{rCl}
			\partial \Delta\tau_{t,r}/\partial \vect{q} &=& ( \vect{u}(\theta_{\text{R},1,1}) - \vect{u}(\theta_{\text{R},t,r}) )/c\\
			\partial \Delta\tau_{t,r}/\partial \alpha_{\text{T}} &=& \big( \vect{u}_{\perp}\trans(\theta_{\text{T},1,1})(\vect{p}_{\text{T},1} - \vect{p}_{\text{T}})\nonumber\\
			&& - \vect{u}_{\perp}\trans(\theta_{\text{T},t,r})(\vect{p}_{\text{T},t} - \vect{p}_{\text{T}}) \big)/c\\
			\partial \tilde{\theta}_{\text{R},t,r}/\partial \vect{q} &=& \vect{u}_{\perp}(\theta_{\text{R},t,r})/d_{t,r}\\
			\partial \tilde{\theta}_{\text{R},t,r}/\partial \alpha_{\text{T}} &=& \vect{u}\trans(\theta_{\text{T},t,r}) (\vect{p}_{\text{T},t} - \vect{p}_{\text{T}})/d_{t,r}.
			\IEEEeqnarraynumspace
		\end{IEEEeqnarray}
		
		Following~\cite{SW07}, we use the notion of equivalent FIM (EFIM) to describe the available information on the relative position and orientation parameters. Splitting $\vect{T}$ as 
		$
			\vect{T} = \begin{bmatrix}
			\vect{T}_{\text{po}}\trans & \vect{T}_{\text{np}}\trans
			\end{bmatrix}\trans,
		$
		with $\vect{T}_{\text{po}}$ consisting of the first three rows of $\vect{T}$ corresponding to the position and orientation parameters and $\vect{T}_{\text{np}}$ the rest,
		the EFIM for the position and orientation parameters is
		\begin{IEEEeqnarray}{rCl}
		\vect{J}_{\text{po}} &=& \vect{T}_{\text{po}} \vect{J}_{\vect{\phi}} \vect{T}_{\text{po}}\trans - \vect{T}_{\text{po}} \vect{J}_{\vect{\phi}} \vect{T}_{\text{np}}\trans \left(\vect{T}_{\text{np}} \vect{J}_{\vect{\phi}}\vect{T}_{\text{np}}\trans\right)^{-1}\vect{T}_{\text{np}} \vect{J}_{\vect{\phi}} \vect{T}_{\text{po}}\trans,
		\IEEEeqnarraynumspace
		\end{IEEEeqnarray}
		where the second term in the right-hand side describes the information loss due to the uncertainty about the rest of the parameters in $\tilde{\vect{\phi}}$.
		
		After some algebraic manipulations, we can show that 
		\begin{IEEEeqnarray}{rCl}
			\vect{J}_{\text{po}} \hspace*{-0.05cm}&=&\hspace*{-0.05cm} \sum_{t=1}^{K_{\text{T}}}\hspace*{-0.04cm} \sum_{r=1}^{K_{\text{R}}}\hspace*{-0.03cm} g_{t,r} \hspace*{-0.06cm}\left( \hspace*{-0.05cm}\frac{\beta_t^2}{c^2}\vect{v}_{\tau,t,r}\vect{v}_{\tau,t,r}\trans \hspace*{-0.04cm}+ \hspace*{-0.04cm}\frac{\omega_{\text{c}}^2 S_r(\tilde{\theta}_{\text{R},t,r})}{c^2d_{t,r}^2}  \vect{v}_{\theta,t,r} \hspace*{-0.02cm}\vect{v}_{\theta,t,r}\trans \hspace*{-0.06cm}\right)\nonumber\\
			&& - \frac{1}{\sum_{t=1}^{K_{\text{T}}}\sum_{r=1}^{K_{\text{R}}} g_{t,r}\beta_t^2/c^2} \vect{v}_{\tau}\vect{v}_{\tau}\trans,
			\IEEEeqnarraynumspace
			\label{eq:main result AOA and TDOA}
		\end{IEEEeqnarray}
		where 
		\begin{IEEEeqnarray}{rCl}
			\vect{v}_{\tau,t,r} &=& \begin{bmatrix}
				\vect{u}\trans(\theta_{\text{R},t,r}) & \vect{u}_{\perp}\trans(\theta_{\text{T},t,r})(\vect{p}_{\text{T},t} - \vect{p}_{\text{T}})
			\end{bmatrix}\trans\label{eq:main result - TDOA vector}\\
			\vect{v}_{\theta,t,r} &=& \begin{bmatrix}
				\vect{u}_{\perp}\trans(\theta_{\text{R},t,r}) & \vect{u}\trans(\theta_{\text{T},t,r})(\vect{p}_{\text{T},t} - \vect{p}_{\text{T}})
			\end{bmatrix}\trans\label{eq:main result - AOA vector}\\
			\vect{v}_{\tau} &=& \sum_{t=1}^{K_{\text{T}}}\sum_{r=1}^{K_{\text{R}}} g_{t,r}\beta_t^2/c^2 \vect{v}_{\tau,t,r}\\
			g_{t,r} &=& 2 N_{\text{R},r}N_{\text{B}} P_{\text{T}} \gamma_t \left|h_{t,r}\right|^2/\sigma_{\eta,r}^2,
		\end{IEEEeqnarray}
		and 
		\begin{IEEEeqnarray}{rCl}
			\beta_t &=& \sqrt{\sum\nolimits_{p\in\mathcal{P}_t}\gamma_{t,p} \omega_p^2 - \left(\sum\nolimits_{p\in\mathcal{P}_t}\gamma_{t,p} \omega_p\right)^2}
			\IEEEeqnarraynumspace
		\end{IEEEeqnarray}
		is the \emph{effective baseband bandwidth} of the signal from the $t$-th Tx array, which is the multi-carrier counterpart of the corresponding quantity defined in \cite{HSZ+16} for single-carrier systems. Also
		\begin{IEEEeqnarray}{rCl}
			S_r(\tilde{\theta}_{\text{R},t,r}) &=& \frac{1}{N_{\text{R},r}} \sum_{i=1}^{N_{\text{R},r}}\left(d_{\text{R},r,i}\vect{u}_{\perp}\trans\left(\psi_{\text{R},r,i}\right)\vect{u}\left(\tilde{\theta}_{\text{R},t,r}\right)\right)^2
			\IEEEeqnarraynumspace
		\end{IEEEeqnarray}
		is the \emph{squared array aperture function} (SAAF) of the $r$-th Rx array, defined in~\cite{HSZ+16}, with the array's centroid chosen as the array's reference point in our case. 
		
		We note that the result for $\vect{J}_{\text{po}}$ is independent of the assumption that $\tau_{1,1} = \min_{t,r} \tau_{t,r}$. 
		The first and second term in the parenthesis in \eqref{eq:main result AOA and TDOA} account for the position and orientation information gain from TDOA and AOA measurements, respectively. We can see in \eqref{eq:main result - TDOA vector} and \eqref{eq:main result - AOA vector} that for each path the position information gain from the TDOA measurement is in the direction of the AOA ($\vect{u}(\theta_{\text{R},t,r})$) and from the AOA measurement in the orthogonal direction ($\vect{u}_{\perp}(\theta_{\text{R},t,r})$). The orientation information gain is determined by the projection of the array's displacement vector $\vect{p}_{\text{T},t} - \vect{p}_{\text{T}}$ on the direction of the AOA for the AOA measurements and its orthogonal direction for the TDOA measurements. 
		There is also information loss due to coupling of the TDOA measurements, which are all taken w.r.t. a common reference. 
		As shown in the last term in \eqref{eq:main result AOA and TDOA}, the direction of the information loss is the weighted average of the directions of information from TDOA measurements. To quantify the position information loss due to the unknown orientation, the notion of EFIM can be applied again to \eqref{eq:main result AOA and TDOA} to obtain the position EFIM. 
		From \eqref{eq:main result AOA and TDOA} we conclude that at least two LOS links are required to have a non-singular $\vect{J}_{\text{po}}$, so that position and orientation estimation is possible. 
		With only one LOS link, position information is obtained solely from AOA estimation, as there is no TDOA and, thus, positioning is not possible.
		
		When $N_{\text{R},r}=1, \; r=1,\ldots, K_{\text{R}}$, although we have in total $K_{\text{R}}$ Rx antenna elements, no angle information can be extracted, according to \eqref{eq:main result AOA and TDOA}, as $S_r(\tilde{\theta}_{\text{R},t,r})=0 \quad \forall r$. The reason is that each link has a different complex gain with \emph{unknown phase}; therefore, the carrier phase difference, which is normally used to extract angle information, cannot be exploited in this case. This is an additional difference from \cite{HSZ+16}, apart from the use of TDOA measurements and the transmitter localization (position and orientation estimation).
		
		In the context of V2V positioning, different requirements for the lateral and longitudinal positioning accuracy are considered.
		Hence, the measures we use to characterize the achievable position estimation accuracy are the lateral positioning error bound $\text{PEB}_{\text{lat}}$ and the longitudinal positioning error bound $\text{PEB}_{\text{lon}}$, defined as
		\begin{IEEEeqnarray}{rCl}
			\text{PEB}_{\text{lat}}&=&\sqrt{\left[\vect{J}_{\text{po}}^{-1}\right]_{1,1}}\label{eq:PEB_lat}\\
			\text{PEB}_{\text{lon}}&=&\sqrt{\left[\vect{J}_{\text{po}}^{-1}\right]_{2,2}}\label{eq:PEB_lon}.
		\end{IEEEeqnarray} 
		
	\subsection{Using only AOA measurements}
		Since in this case we  aim to determine the achievable position and orientation estimation accuracy using only AOA measurements, we now define a different parameter vector $\tilde{\vect{\phi}}$, which includes the TDOAs as nuisance parameters:
		\begin{IEEEeqnarray}{rCl}
			\tilde{\vect{\phi}} &=& [\vect{q}\trans, \alpha_{\text{T}}, \tau_{\text{s}}, h_{1,1, \Re}, h_{1,1, \Im}, \Delta\tau_{1,2}, h_{1,2, \Re}, h_{1,2, \Im}, \ldots, \nonumber\\
			&& \Delta\tau_{K_{\text{T}},K_{\text{R}}}, h_{K_{\text{T}},K_{\text{R}}, \Re}, h_{K_{\text{T}},K_{\text{R}}, \Im}
			]\trans\in\mathbb{R}^{3 + 3K_{\text{T}} K_{\text{R}}}\label{eq:position parameter vector AOA}.
			\IEEEeqnarraynumspace
		\end{IEEEeqnarray}
		Using similar steps as before, we can show that the position and orientation EFIM can be expressed as
		\begin{IEEEeqnarray}{rCl}
			\vect{J}_{\text{po}} &=& \sum_{t=1}^{K_{\text{T}}}\sum_{r=1}^{K_{\text{R}}} g_{t,r} \frac{\omega_{\text{c}}^2 S_r(\tilde{\theta}_{\text{R},t,r})}{c^2d_{t,r}^2}  \vect{v}_{\theta,t,r} \vect{v}_{\theta,t,r}\trans\label{eq:main result AOA only},
			\IEEEeqnarraynumspace
		\end{IEEEeqnarray}
		which, as expected, corresponds to the second term in \eqref{eq:main result AOA and TDOA}. In contrast to \eqref{eq:main result AOA and TDOA}, in this case at least three LOS links are required to have a non-singular $\vect{J}_{\text{po}}$.

\section{Numerical results}
\begin{figure}
		\centering
		\subfloat[Overtaking scenario.]{
			\begin{adjustbox}{scale=0.88}\includegraphics[]{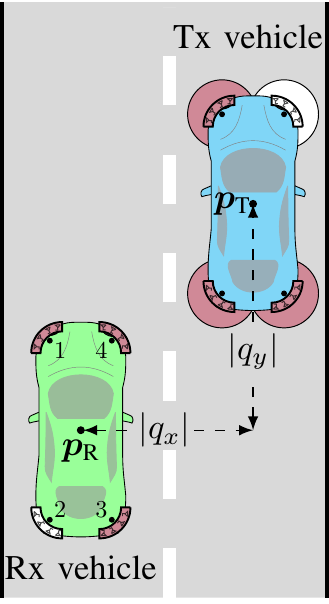}\end{adjustbox}
		\label{fig:overtaking_scenario}}\quad
		\subfloat[Platooning scenario.]{
			\begin{adjustbox}{scale=0.88}\includegraphics[]{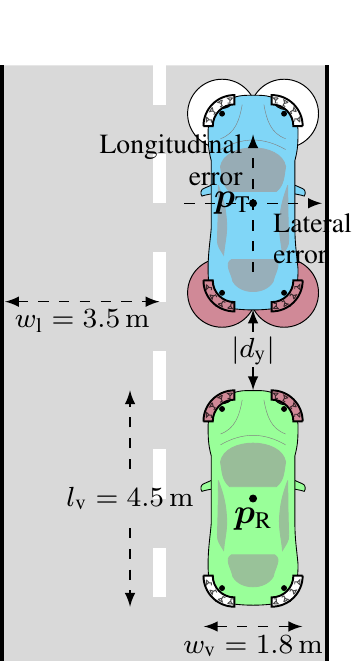}\end{adjustbox}
			\label{fig:platooning_scenario}}
		\caption{Overtaking and platooning scenarios. Arrays with at least one LOS link are shown darker.}
		\label{fig:scenarios}
	\end{figure}

	In this section we evaluate the derived bounds for some V2V scenarios of interest, namely overtaking and platooning, which are shown in Fig. \ref{fig:scenarios}. We compare the bounds with the positioning accuracy requirements defined in \cite{TS22.186}, where it is stated that the lateral and longitudinal position errors (see Fig. \ref{fig:platooning_scenario}) should be less than \SI{0.1}{\meter} and \SI{0.5}{\meter}, respectively. The Tx and Rx vehicles, with length $l_{\text{v}} = \SI{4.5}{\meter}$ and width $w_{\text{v}} = \SI{1.8}{\meter}$, have $K_{\text{T}} = K_{\text{R}} = 4$ \emph{conformal} arrays, whose centroids are located at the vehicles' corners. The arrays are designed as appropriate quarters of a uniform circular array with $\lambda_{\text{c}}/2$-spaced elements, where $\lambda_{\text{c}}$ is the carrier wavelength, as shown in Fig. \ref{fig:scenarios}. 
	The fixed beamforming vector for each Tx array is chosen so that the signals are transmitted omnidirectionally in the \ang{270}-sector that is not blocked by the vehicle (see Fig. \ref{fig:scenarios}). Hence, all potential LOS links are excited, without the need of a beam-scanning process. Similarly, each of the Rx arrays can receive signals in an angular range of \ang{270}.
	Thus, the links between some Tx-Rx array pairs might be blocked. In Fig. \ref{fig:scenarios} the arrays that have at least one LOS link are shown darker. The width of the lane is assumed to be $w_{\text{l}}=\SI{3.5}{\meter}$.
	The channel gain $h_{t,r}'$ of each link is computed assuming free space propagation for the respective wavelength.
	We set $N_{\text{B}}=1, N=2048$ and $\mathcal{P}=\{-600,\ldots,-1,1\ldots,600\}$, with the used subcarriers uniformly distributed to the Tx arrays in an interleaved manner. Also, the power is uniformly allocated among Tx arrays and subcarriers, i.e. $\gamma_t = 1/4,\;\gamma_{t,p}=1/|\mathcal{P}_t|, \forall t,p$. For the rest of the system parameters we consider two configurations: 
	\begin{enumerate}[(i)]
		\item $f_{\text{c}}=\SI{3.5}{\GHz}$, $\Delta f = \SI{60}{kHz}$ and $N_{\text{R},r}=4,\forall r$; 
		\item $f_{\text{c}}=\SI{28}{\GHz}$, $\Delta f = \SI{240}{kHz}$ and $N_{\text{R},r}=25,\forall r$,
	\end{enumerate}
	where $\Delta f$ denotes the subcarrier spacing. The transmit power $P_{\text{T}}$ and the per-antenna noise variance $\sigma_{\eta,r}^2$ are set so that, when the vehicles are next to each other in neighboring lanes ($q_x=\SI{-3.5}{\meter, q_y=\SI{0}{m}}$), the receive SNR after Rx beamforming $g_{t,r}/|\mathcal{P}_t|$ for the Tx-Rx array pair with the shortest distance is \SI{36}{\decibel} for the \SI{3.5}{\GHz} configuration and \SI{30}{\decibel} for the \SI{28}{\GHz} configuration. 
	The number of antenna elements for the two configurations has been chosen so that the receiving arrays have (approximately) the same size, assuming the inter-element spacing is $\lambda_{\text{c}}/2$.
	Numbering the arrays as shown for the Rx vehicle in Fig. \ref{fig:overtaking_scenario}, the antenna elements of the $r$-th array are located at $\vect{p}_{\text{R},r,i} = \vect{p}_{\text{R},r} + \rho\vect{u}\left(\delta_i + \frac{\pi}{2}(r - 1) + \alpha_{\text{R}}\right) - \bar{\vect{p}}$, where 
	\begin{IEEEeqnarray}{rCl}
		\rho &=& \lambda_{\text{c}}/\left(4 \sin\left(\pi\middle/\left(4\left(N_{\text{R}, r} - 1\right)\right)\right)\right)\nonumber\\
		\delta_i &=& \pi\left(i - 1\right)/\left(2\left(N_{\text{R}, r} - 1\right)\right)\nonumber\\
		\bar{\vect{p}} &=& \rho\sum\nolimits_{i=1}^{N_{\text{R}, r}} \vect{u}(\delta_i)/N_{\text{R}, r}\nonumber.
	\end{IEEEeqnarray}
	In the following, we discuss the two aforementioned scenarios.
	
	\subsection{Overtaking Scenario}
\begin{figure}
		\centering
		\subfloat[Lateral position error bound.]{
			\begin{adjustbox}{scale=0.94}\includegraphics[]{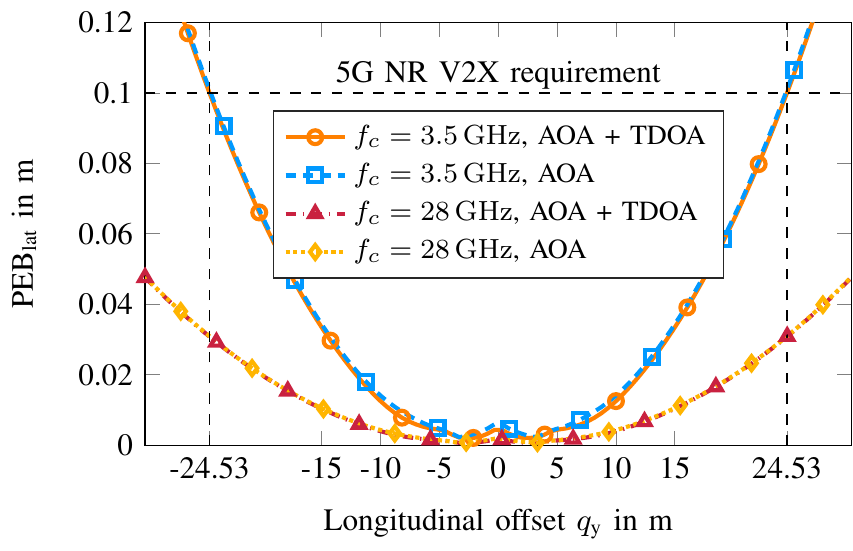}\end{adjustbox}
		}\quad
		\subfloat[Longitudinal position error bound.]{
			\begin{adjustbox}{scale=0.94}\includegraphics[]{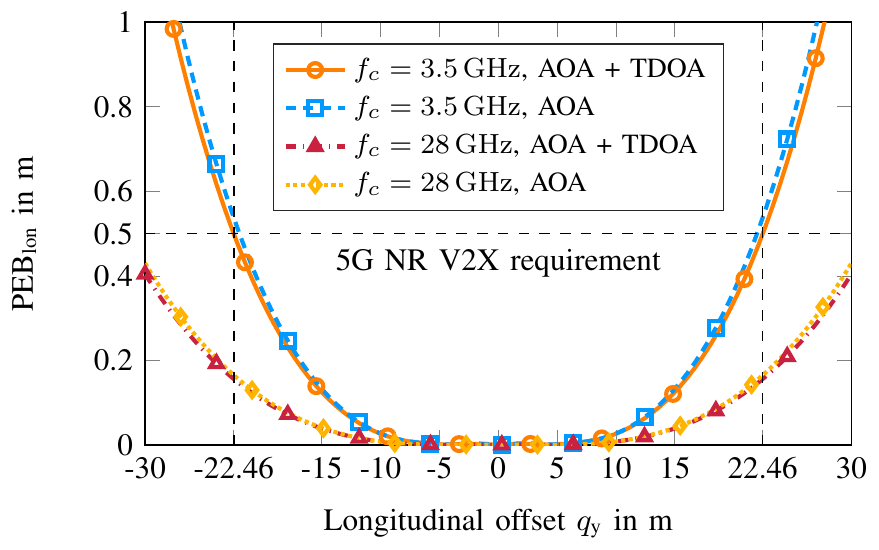}\end{adjustbox}
		}
		\caption{Error bounds for the overtaking scenario.}
		\label{fig:overtaking_CRB}
	\end{figure}

	The overtaking scenario is shown in Fig. \ref{fig:scenarios}a. The lateral offset of the Rx vehicle's center w.r.t. to the Tx vehicle's center is constant and equal to $q_{\text{x}}=\SI{-3.5}{\meter}$ (lane width) and a range of longitudinal offsets $q_{\text{y}}$ from \SI{-30}{\meter} to \SI{30}{\meter} is considered. For all considered relative positions in this scenario, each vehicle has three arrays with at least one LOS link to the other vehicle, except for $q_{\text{y}}=\SI{0}{m}$ (where the vehicles are next to each other); then, each vehicle has two arrays with at least one LOS link. In Fig. \ref{fig:overtaking_CRB} we plot the lateral and the longitudinal positioning errors $\text{PEB}_{\text{lat}}$ \eqref{eq:PEB_lat} and $\text{PEB}_{\text{lon}}$ \eqref{eq:PEB_lon} as functions of the longitudinal offset $q_{\text{y}}$, when both AOA and TDOA \eqref{eq:main result AOA and TDOA} or only AOA measurements \eqref{eq:main result AOA only} are used, for the two aforementioned configurations.
	For both configurations, we observe that TDOA measurements do not provide much additional positioning information compared to that provided by the AOA measurements. Despite the lower receive SNR, the \SI{28}{\GHz} configuration has better positioning accuracy than the \SI{3.5}{\GHz} configuration, with the higher angular resolution offered by the higher number of antennas being the key factor for its superiority. 
	We stress that, as we can see in \eqref{eq:main result AOA and TDOA} and \eqref{eq:main result AOA only}, the bandwidth does not impact the angle information. 
	While with the \SI{28}{\GHz} configuration the 5G V2X positioning requirements are always met for the depicted vehicle distances, the \SI{3.5}{\GHz} configuration achieves the lateral positioning accuracy requirement for longitudinal offsets $|q_{\text{y}}| < \SI{24.53}{\meter}$ and the longitudinal requirement for $|q_{\text{y}}|<\SI{22.46}{\meter}$, when both AOA and TDOA measurements are used. For the case when solely AOA measurements are used, the corresponding values are only slightly lower.
	
	\subsection{Platooning Scenario}
\begin{figure}
		\centering
		\subfloat[Lateral position error bound.]{
			\begin{adjustbox}{scale=0.93}\includegraphics[]{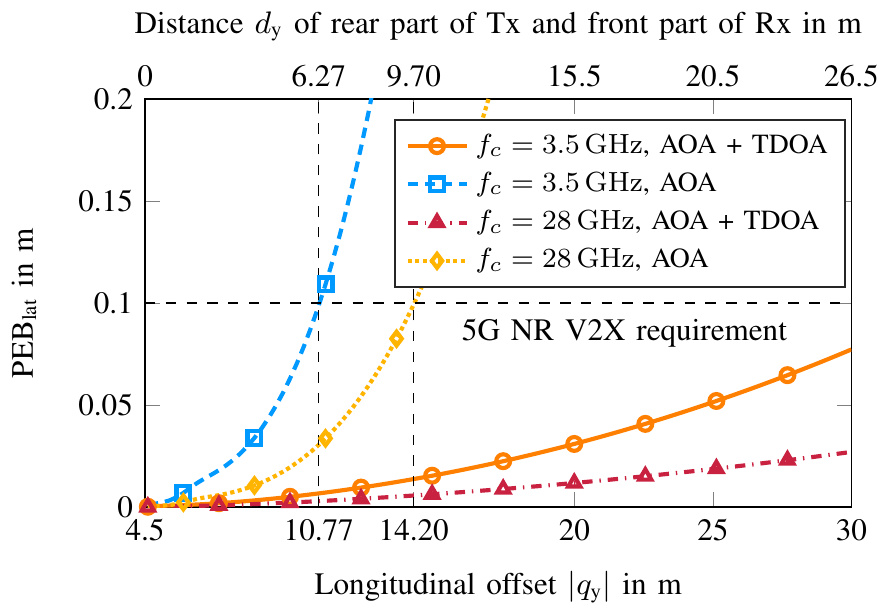}\end{adjustbox}
		}\quad
		\subfloat[Longitudinal position error bound.]{
			\begin{adjustbox}{scale=0.93}\includegraphics[]{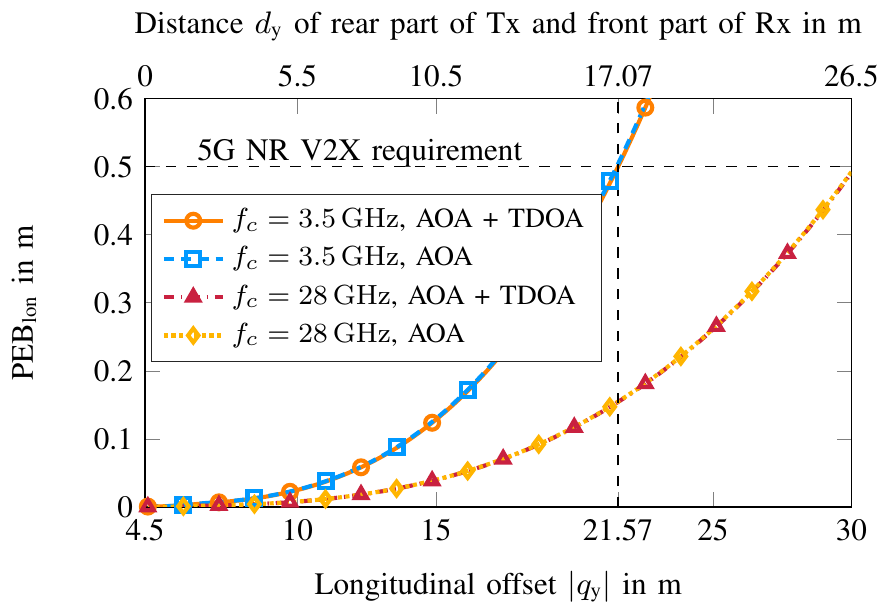}\end{adjustbox}
		}
		\caption{Error bounds for the platooning scenario.}
		\label{fig:platooning_CRB}
	\end{figure}

	In the platooning scenario the vehicles are vertically aligned, i.e. $q_{\text{x}} = \SI{0}{m}$. Consequently, as seen in Fig. \ref{fig:scenarios}, only the two rear panels of the Tx vehicle and the two front panels of the Rx vehicle have an active communication link. We consider only negative longitudinal offsets $\SI{-30}{\meter} < q_{\text{y}} \leq \SI{-4.5}{m}$ and in Fig. \ref{fig:platooning_CRB} we plot $\text{PEB}_{\text{lat}}$ and $\text{PEB}_{\text{lon}}$ as functions of $|q_{\text{y}}|$ (lower horizontal axis) and of $d_{\text{y}} = |q_{\text{y}}| - \SI{4.5}{\meter}$ (upper horizontal axis), which is the distance between the rear part of the Tx vehicle and the front part of the Rx vehicle (see Fig. \ref{fig:platooning_scenario}).
	The \SI{28}{\GHz} configuration provides again better lateral and longitudinal positioning accuracy. As we can see in Fig. \ref{fig:platooning_CRB}b, similar to the overtaking scenario, the use of TDOA measurements in addition to the AOA measurements has a very small impact on the longitudinal positioning accuracy, with the 3GPP longitudinal requirement being satisfied for $d_{\text{y}} < \SI{17.07}{\meter}$ for the \SI{3.5}{\GHz} configuration and for all considered distances for the \SI{28}{\GHz} configuration. On the other hand, we see that the TDOA measurements strongly influence the lateral position error bound. When AOA and TDOA are used, the lateral requirement is satisfied for all considered distances for both configurations. However, when only AOA measurements are used, the lateral requirement can only be met for $d_{\text{y}}<\SI{6.27}{\meter}$ for the \SI{3.5}{\GHz} configuration and for $d_{\text{y}}<\SI{9.70}{\meter}$ for the \SI{28}{\GHz} configuration.
	Numerical results, not included here due to space constraints, show that i) when the orientation $\alpha_{\text{T}}$ of  the Tx vehicle is known, the TDOA measurements do not significantly improve the lateral positioning accuracy; ii) when $\alpha_{\text{T}}$ is unknown the TDOA measurements substantially increase the Fisher information on $\alpha_{\text{T}}$. These two observations explain the reason why TDOA measurements are important for this scenario.
	
\section{Conclusion}
	We derived the Fisher information and the CRB on the relative position and orientation estimation error for vehicles equipped with multiple antenna arrays. We evaluated the bounds for different configurations and relevant scenarios, computing the distances, for which the 5G V2X requirements are met. Our results show that the configuration with the \SI{28}{\GHz} carrier frequency can provide better positioning accuracy due to its higher angular resolution. We also showed that AOA measurements are much more important than TDOA measurements for the lateral position error in the overtaking scenario and for the longitudinal position error in both scenarios. Nevertheless, TDOA measurements can drastically improve the lateral positioning accuracy in the platooning scenario, as they provide valuable Fisher information on the orientation of the Tx vehicle. The derived bounds can be used for the optimization of the position of the arrays and their elements to meet specific lateral and longitudinal positioning accuracy requirements.
	Further work on the topic includes the extension to time-varying multipath channels.

\section*{Acknowledgment}
	Part of this work has been performed in the framework of the H2020 project 5GCAR co-funded by the EU. The authors would like to acknowledge the contributions of their colleagues from 5GCAR although the views expressed are those of the authors and do not necessarily represent the views of the 5GCAR project.

\bibliographystyle{IEEEtran}
\bibliography{IEEEabrv,Multi-Array_5G_V2V_Relative_Positioning_Performance_Bounds}

\begin{thebibliography}{10}
\providecommand{\url}[1]{#1}
\csname url@samestyle\endcsname
\providecommand{\newblock}{\relax}
\providecommand{\bibinfo}[2]{#2}
\providecommand{\BIBentrySTDinterwordspacing}{\spaceskip=0pt\relax}
\providecommand{\BIBentryALTinterwordstretchfactor}{4}
\providecommand{\BIBentryALTinterwordspacing}{\spaceskip=\fontdimen2\font plus
\BIBentryALTinterwordstretchfactor\fontdimen3\font minus
  \fontdimen4\font\relax}
\providecommand{\BIBforeignlanguage}[2]{{%
\expandafter\ifx\csname l@#1\endcsname\relax
\typeout{** WARNING: IEEEtran.bst: No hyphenation pattern has been}%
\typeout{** loaded for the language `#1'. Using the pattern for}%
\typeout{** the default language instead.}%
\else
\language=\csname l@#1\endcsname
\fi
#2}}
\providecommand{\BIBdecl}{\relax}
\BIBdecl

\bibitem{HWC+15}
A.~Hakkarainen, J.~Werner, M.~Costa, K.~Leppanen, and M.~Valkama,
  ``High-efficiency device localization in 5{G} ultra-dense networks: Prospects
  and enabling technologies,'' in \emph{Proc. IEEE 82th Veh. Technol. Conf.
  (VTC-Fall), \normalfont Boston, MA}, Sep. 2015, pp. 1--5.

\bibitem{SGD+18}
A.~Shahmansoori, G.~E. Garcia, G.~Destino, G.~Seco-Granados, and H.~Wymeersch,
  ``Position and orientation estimation through millimeter-wave {MIMO} in 5{G}
  systems,'' \emph{{IEEE} Trans. Wireless Commun.}, vol.~17, no.~3, pp.
  1822--1835, Mar. 2018.

\bibitem{WML+16}
K.~Witrisal \emph{et~al.}, ``High-accuracy localization for assisted living:
  5{G} systems will turn multipath channels from foe to friend,'' \emph{{IEEE}
  Signal Process. Mag.}, vol.~33, no.~2, pp. 59--70, Mar. 2016.

\bibitem{WHK+16}
K.~Witrisal, S.~Hinteregger, J.~Kulmer, E.~Leitinger, and P.~Meissner,
  ``High-accuracy positioning for indoor applications: {RFID}, {UWB}, 5{G}, and
  beyond,'' in \emph{Proc. IEEE 10th Int. Conf. RFID}, May 2016, pp. 1--7.

\bibitem{WSD+17}
H.~Wymeersch, G.~Seco-Granados, G.~Destino, D.~Dardari, and F.~Tufvesson,
  ``5{G} mm{W}ave positioning for vehicular networks,'' \emph{IEEE Wireless
  Commun.}, vol.~24, no.~6, pp. 80--86, Dec 2017.

\bibitem{HKC+18}
K.~{Han} \emph{et~al.}, ``Sensing hidden vehicles by exploiting multi-path
  {V2V} transmission,'' \emph{ArXiv e-prints}, Apr. 2018.

\bibitem{HSZ+16}
Y.~Han, Y.~Shen, X.~P. Zhang, M.~Z. Win, and H.~Meng, ``Performance limits and
  geometric properties of array localization,'' \emph{{IEEE} Trans. Inf.
  Theory}, vol.~62, no.~2, pp. 1054--1075, Feb. 2016.

\bibitem{SGD+15}
A.~Shahmansoori, G.~E. Garcia, G.~Destino, G.~Seco-Granados, and H.~Wymeersch,
  ``5{G} position and orientation estimation through millimeter wave {MIMO},''
  in \emph{Proc. IEEE Globecom Workshops (GC Wkshps), \normalfont San Diego,
  CA}, Dec. 2015, pp. 1--6.

\bibitem{GGD17}
A.~Guerra, F.~Guidi, and D.~Dardari, ``Single anchor localization and
  orientation performance limits using massive arrays: {MIMO} vs.
  beamforming,'' \emph{ArXiv e-prints}, Feb. 2017.

\bibitem{AZA+17}
Z.~Abu{-}Shaban, X.~Zhou, T.~D. Abhayapala, G.~Seco{-}Granados, and
  H.~Wymeersch, ``Error bounds for uplink and downlink {3D} localization in
  {5G} {mmWave} systems,'' \emph{ArXiv e-prints}, Apr. 2017.

\bibitem{MWB+17}
R.~{Mendrzik}, H.~{Wymeersch}, G.~{Bauch}, and Z.~{Abu-Shaban}, ``Harnessing
  {NLOS} components for position and orientation estimation in 5{G} mm{W}ave
  {MIMO},'' \emph{ArXiv e-prints}, Dec. 2017.

\bibitem{DDG+17}
D.~Dardari \emph{et~al.}, ``High-accuracy tracking using ultrawideband signals
  for enhanced safety of cyclists,'' \emph{Mobile Information Systems}, vol.
  2017, Article ID 8149348, 13 pages, 2017.

\bibitem{TS22.186}
{3rd Generation Partnership Project (3GPP)}, ``{Technical Specification Group
  Services and System Aspects; Enhancement of 3GPP support for V2X scenarios;
  Stage 1 (Release 15)},'' TS22.186 V15.2.0, Sep. 2017.

\bibitem{Kay93}
S.~M. Kay, \emph{Fundamentals of Statistical Signal Processing: Estimation
  Theory}.\hskip 1em plus 0.5em minus 0.4em\relax Upper Saddle River, NJ, USA:
  Prentice-Hall, Inc., 1993.

\bibitem{LC98}
E.~L. Lehmann and G.~Casella, \emph{{Theory of Point Estimation (Springer Texts
  in Statistics)}}, 2nd~ed.\hskip 1em plus 0.5em minus 0.4em\relax Springer New
  York, Aug. 1998.

\bibitem{SW07}
Y.~Shen and M.~Z. Win, ``Fundamental limits of wideband localization accuracy
  via fisher information,'' in \emph{Proc. IEEE Wireless Commun. Netw. Conf.
  (WCNC), \normalfont Kowloon, Hong Kong}, Mar. 2007, pp. 3046--3051.

\end{thebibliography}

\end{document}